\documentclass[aip,jmp,amsmath,amssymb,reprint,]{revtex4-1}

\usepackage[pdftex]{graphicx}
\usepackage{dcolumn}
\usepackage{bm}

\begin{document}


\title{Non-resonant feeding of photonic crystal nanocavity modes by quantum dots} 

\author{A. Laucht}
\author{N. Hauke}
\author{A. Neumann}
\author{T. G\"unthner}
\author{F. Hofbauer}
\author{A. Mohtashami}
\author{K. M\"uller}
\author{G. B\"ohm}
\author{M. Bichler}
\author{M.-C. Amann}
\author{M. Kaniber}
\author{J. J. Finley}%
\email{finley@wsi.tum.de}
\affiliation{Walter Schottky Institut, Technische Universit\"at M\"unchen, Am Coulombwall 3, 85748 Garching, Germany}%

\date{\today}

\begin{abstract}
We experimentally probe the non-resonant feeding of photons into the optical mode of a two dimensional photonic crystal nanocavity from the discrete emission from a quantum dot. For a strongly coupled system of a single exciton and the cavity mode, we track the detuning-dependent photoluminescence intensity of the polariton peaks at different lattice temperatures. At low temperatures we observe a clear asymmetry in the emission intensity depending on whether the exciton is at higher or lower energy than the cavity mode. At high temperatures this asymmetry vanishes when the probabilities to emit or absorb a phonon become similar.
For a different dot-cavity system where the cavity mode is detuned by $\Delta E>5$~meV to lower energy than the single exciton transitions emission from the mode remains correlated with the quantum dot as demonstrated unambiguously by cross-correlation photon counting experiments. By monitoring the temporal evolution of the photoluminescence spectrum, we show that feeding of photons into the mode occurs from multi-exciton transitions. We observe a clear anti-correlation of the mode and single exciton emission; the mode emission quenches as the population in the system reduces towards the single exciton level whilst the intensity of the mode emission tracks the multi-exciton transitions.
\end{abstract}


\maketitle 

%
%
The atom like properties of semiconductor quantum dots (QDs) make them ideal candidates for cavity quantum electrodynamics (cQED) experiments in the solid state.~\cite{Vahala03} QD-cavity systems are used for many different applications including the efficient and deterministic generation of indistinguishable photons,~\cite{Santori02} devices that exploit single photon quantum non-linearities,~\cite{Englund07} and ultra low threshold nanolasers.~\cite{Strauf06} Whilst these systems exhibit many effects already known from atomic cQED experiments, the solid-state environment leads to a number of significant deviations from this model system. Photoluminescence studies of photonic crystal defect nanocavities containing a few QD emitters typically reveal intense emission from the cavity mode even though dot and cavity are spectrally detuned. For small dot-cavity detunings ($\left|E\right|<5$~meV), acoustic phonon mediated dot - cavity coupling has been shown to feed this cavity emission.~\cite{Hohenester09, Suffcynski09, Hohenester10, Ota09} However, even when the discrete QD transitions and the mode are strongly detuned from one another by $\left|\Delta E\right|>5$~meV a mechanism exists by which the QDs non-resonantly couple to the cavity mode.~\cite{Hennessy07, Press07, Kaniber08b, Winger09} For these large detunings, acoustic phonon mediated coupling becomes ineffective and the cavity mode emission stems from optical transitions between higher excited multi-exciton states and an energetically lower quasi-continuum of multi-exciton states.~\cite{Kaniber08b, Winger09, Laucht10b}

%
%
In this paper we review the non-resonant photon feeding phenomenon from a single QD into the cavity mode of a two-dimensional photonic crystal. For a strongly coupled system of a single exciton and the cavity mode, we track the photoluminescence (PL) intensity of the two polariton peaks through the anti-crossing when the detuning is varied by changing the bias voltage applied to our electrically tunable single dot-cavity samples.\cite{Laucht09, Laucht09b} At low temperatures we observe a clear asymmetry in the relative emission intensities; the exciton line is significantly brighter when it is tuned to the low energy side of the cavity mode compared to the same blue detuning. At high temperatures, however, this asymmetry vanishes, indicating that the off-resonant feeding process is phonon mediated. Here, the probability to couple to the cavity mode via a phonon mediated process \cite{Hohenester09, Hohenester10} becomes equallly likely for phonon emission and absorption when $k_b T\gg E_{phonon}$.
For a strongly detuned dot-cavity system where the cavity mode is detuned by $\Delta E\sim 18$~meV to lower energy than the single exciton transition, we show cross-correlation measurements that clearly relate the cavity mode emission to the emission of the QD. Temporally resolved saturation spectroscopy measurements for a cavity mode detuned by $\Delta E\sim 9$~meV to lower energy than the single exciton transition allow us to monitor the temporal evolution of the complete PL spectrum from the QD-cavity system. Whilst the single exciton emission is delayed due to the cascaded emission in the QD,~\cite{Laucht10b, Chauvin09} emission from the cavity mode occurs immediately after the laser pulse is incident on the sample and rapidly quenches as the population in the system reduces towards the single exciton level. Therefore, we conclude that feeding of photons into the mode occurs from multi-exciton transitions for such strong detuning.

%
%
The samples investigated were grown by molecular beam epitaxy and consisted of a $180$~nm thick GaAs slab waveguide grown on a $500$~nm thick Al$_{0.8}$Ga$_{0.2}$As sacrificial layer. For the electrically tunable sample, the slab waveguide was doped to produce a p-i-n photodiode structure.\cite{Hofbauer07, Laucht09} A single layer of In$_{0.5}$Ga$_{0.5}$As QDs ($\sim5$~$\mu$m$^{-2}$) was incorporated at the midpoint of this waveguide. A two-dimensional photonic crystal formed by a triangular array of air holes was realized using a combination of electron-beam lithography and reactive ion etching. Nanocavities were formed by introducing point defects consisting of three missing holes in a row (L3 cavity).~\cite{Akahane03} Within such a cavity, we normally observe emission from $1-3$~QDs. In a final processing step, free standing GaAs membranes were formed by wet etching with hydrofluoric acid.

%
%
For optical characterization the sample was mounted in a liquid He-flow cryostat and cooled to $10-15$~K unless stated otherwise. For the excitation we either used a pulsed Ti-Sapphire laser ($80$~MHz repetition frequency, $2$~ps pulse duration) or a continuous wave Ti-Sapphire laser tuned to a wavelength spectrally in resonance with the wetting layer absorption continuum or a higher energy cavity mode.\cite{Kaniber09} The QD PL was collected using a $100\times$ microscope objective, spectrally analyzed using a $0.5$~m imaging monochromator and detected using a Si-based, liquid nitrogen cooled CCD. For time-resolved spectroscopy we used a silicon avalanche photodiode connected to the side-exit of our monochromator. We obtain a temporal resolution of $\sim150$~ps after deconvolution with the system instrument response function (IRF).

%
%
\begin{figure}[t!]
\includegraphics[width=0.95\columnwidth]{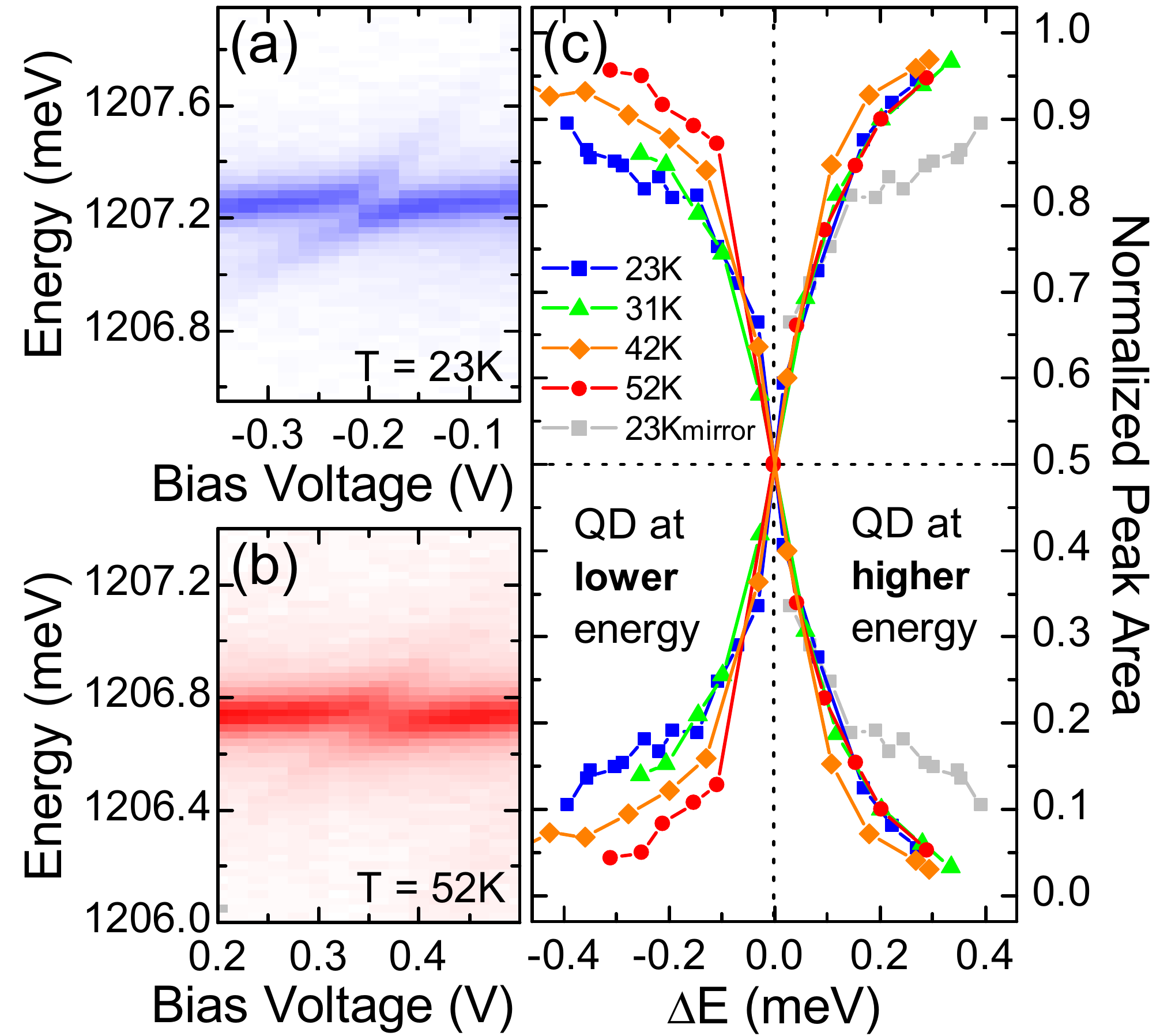}
\caption{\label{figure1} (color online) Photoluminescence spectra of a strongly coupled dot-cavity system as a function of bias voltage (false-color plot) for (a) T=23~K and (b) T=52~K. (c) Extracted, normalized peak area of the two polariton lines for different temperatures as a function of detuning between exciton and cavity mode.}
\end{figure}
In Fig.~\ref{figure1} we present optical investigations performed on a strongly coupled, electrically tunable QD-cavity sample. We show in Fig.\ref{figure1}(a) a false-color plot of the emission spectra for a lattice temperature of $T=23$~K. The exciton is tuned through resonance with the cavity mode as we change the bias voltage due to the quantum confined Stark effect. We observe a clear anticrossing in the PL spectra for $V_{app}=-0.2$~V, with a vacuum Rabi splitting $2g\sim120$~$\mu$eV. In Fig.~\ref{figure1}(b), we plot PL spectra of the same system at a higher temperature of $T=52$~K. At this temperature, pure dephasing due to interaction with acoustic phonons broadens the two polariton peaks and reduces the coherent interaction between exciton and cavity mode.~\cite{Laucht09b} As a result, the anticrossing at resonance is not as clearly visible as for the low temperature measurements. For these two temperatures and two intermediate temperatures ($T=31$~K and $T=42$~K) we extracted the peak area of the two polariton peaks and normalized it to the total luminescence intensity of both polariton peaks. The results of these measurements are plotted in Fig.~\ref{figure1}(c) as a function of detuning from resonance $\Delta E=\sqrt{(E_{peak1}-E_{peak2})^2-(2g)^2}$. The region of the data presented in Fig.\ref{figure1}(c) with higher intensity corresponds to the cavity-like emission peak and region with lower intensity corresponds to the exciton-like emission peak. In resonance ($\Delta E=0$~meV), the exciton and the cavity mode enter the strong coupling regime and an exciton-polariton is formed. Both peaks have equal exciton- and photon-like character and their emission intensity is equal. We now focus the attention to the relative intensities of the exciton- and cavity-like peaks when they are not in resonance. When the QD is at higher energy than the cavity mode ($\Delta E>0$~meV) it can generate a photon in the cavity mode by a phonon mediated emission process.\cite{Hohenester09, Hohenester10} For this case we do not observe a significant difference in intensities as we raise the temperature since a phonon can always be emitted. In strong contrast, when the QD is detuned to lower energy than the cavity mode ($\Delta E<0$~meV) the normalized intensity of the exciton-like polariton branch is clearly highest for the lowest temperatures investigated ($T=23$~K). Upon increasing the temperatures its relative emission intensity decreases, and exhibits the minimum value at the highest temperature investigated ($T=53$~K). These observations can be explained by the strong temperature dependence of the phonon-mediated scattering into the cavity-like polariton branch.~\cite{Hohenester09,Hohenester10} It is only for temperatures such that $k_B T\gg E_{phonon}$, that the phonon absorption rate is sufficiently high that scattering of the exciton into the cavity mode via a phonon absorption process can take place. These observations are fully consistent with recent publications where phonon-mediated, Purcell-enhanced photon feeding of the cavity mode was shown to be active for small QD-cavity detunings.~\cite{Hohenester09, Suffcynski09, Hohenester10, Ota09}

%
%
\begin{figure}[t!]
\includegraphics[width=0.95\columnwidth]{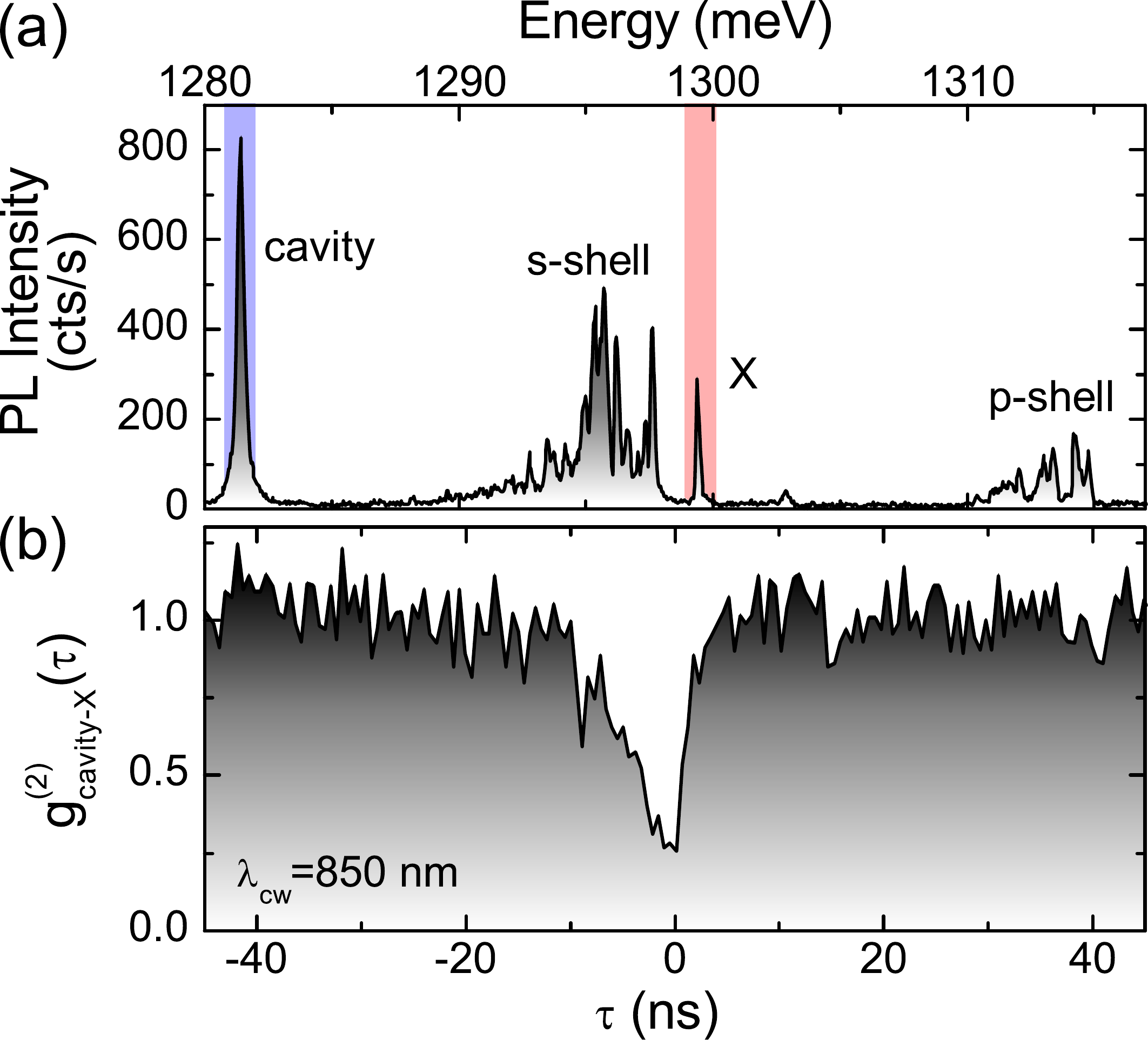}
\caption{\label{figure2} (color online) (a) Photoluminescence spectrum of a QD-cavity system with the cavity mode (marked in blue) detuned to $\Delta E\sim 18$~meV lower energy than the single exciton transition (marked in red). (b) Cross-correlation histogram between the cavity mode and the single exciton. Here, $\tau>0$ corresponds to detection of a photon from the single exciton upon detection of a photon from the cavity mode.}
\end{figure}
Whilst phonon-mediated feeding of the cavity mode is effective only for small detunings ($\left|\Delta E\right|<5$~meV) an additional mechanism must exist that non-resonantly feeds photons into the cavity mode for much larger detunings.~\cite{Hennessy07, Press07, Kaniber08b, Winger09} In Fig.\ref{figure2}(a) we plot the PL spectrum of a different QD-cavity system where the cavity mode is strongly detuned to lower energy than the QD s-shell excitonic emission ($\Delta E\sim15$~meV). Although no discrete QD transitions are present in the spectral vicinity of the cavity mode, pronounced PL emission is still observed from the cavity mode. For this system we conduct photon-cross-correlation measurements between the cavity mode (marked with the blue shaded region on Fig.\ref{figure2}(a)) and a QD single exciton transition (marked with the red shaded region on Fig.\ref{figure2}(a)). The single exciton transition was clearly identified from its linear power dependence. These two emission features are detuned by $\Delta E\sim18$~meV.
In Fig.\ref{figure2}(b) we plot the cross-correlation histogram as a function of the time $\tau$ between two detection events. Here, $\tau>0$ corresponds to detection of a photon from the exciton after detection of a photon from the cavity mode. At zero time delay ($\tau=0$~ns) we observe a dip of $g^{(2)}_{cavity-X}(0)=0.26$ in the histogram. This is a clear signature of strongly anti-correlated photon emission between the cavity mode and the investigated exciton transition. This observation proves that the cavity mode is predominantly fed by this QD.~\cite{Hennessy07,Kaniber08b,Winger09} The asymmtry of the dip is a result of the different lifetimes of the cavity mode and the exciton.\cite{Kaniber08b} Whilst this measurement proves that the origin of the cavity mode emission is indeed due to the QD, it does not allow us to draw any conclusion about the mechanism that mediates such non-resonant coupling.

%
%
\begin{figure}[t!]
\includegraphics[width=0.95\columnwidth]{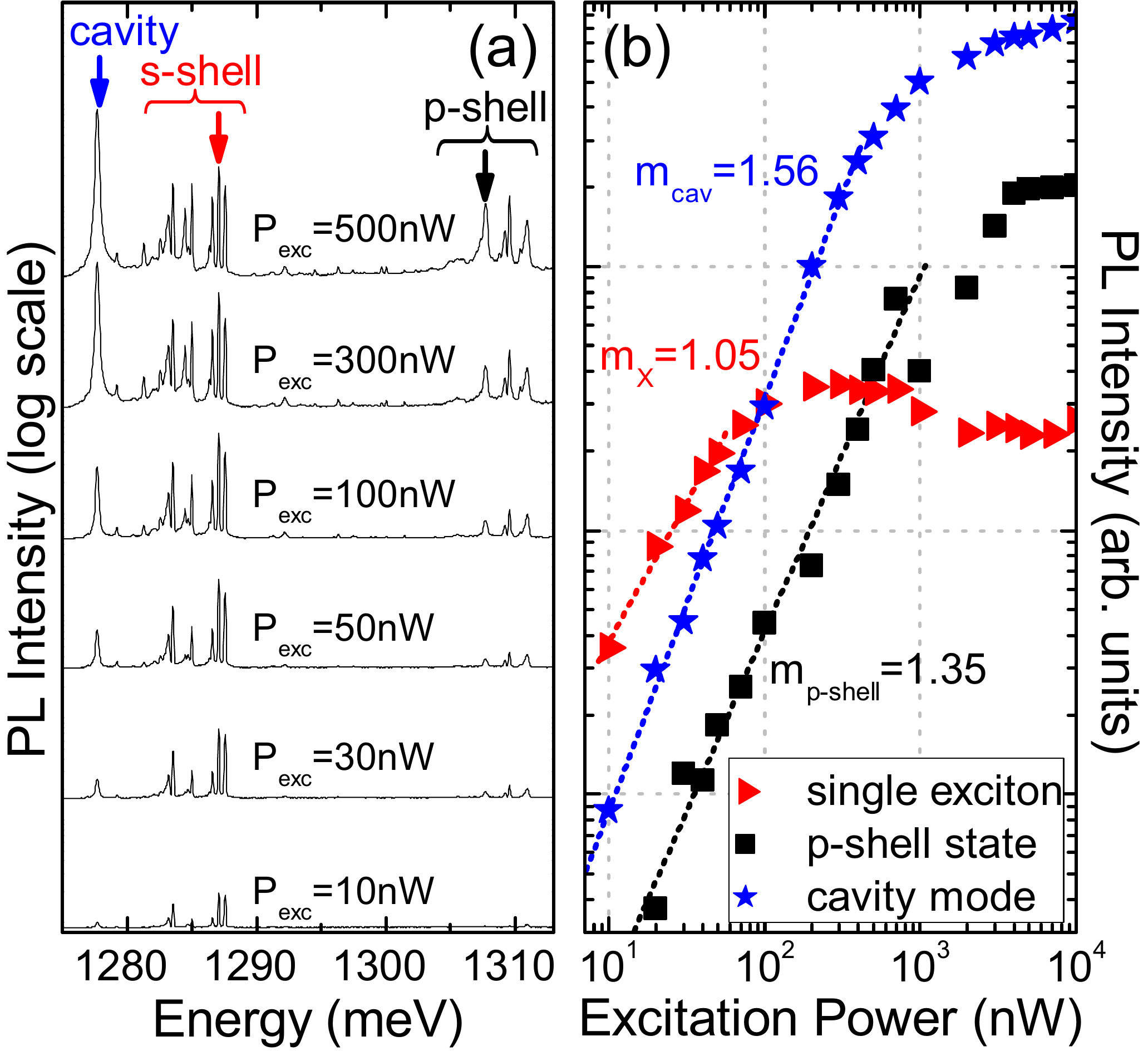}
\caption{\label{figure3} (color online) (a) Power dependent photoluminescence spectra recorded from a QD-cavity system with the cavity mode at lower energy than the QD s-shell. The excitation level increases from bottom to top. (b) Integrated photoluminescence intensity from the cavity mode (blue stars), a selected single exciton transition from the s-shell (red triangles), and a selected p-shell multi-exciton transition (black squares).}
\end{figure}
%
%
\begin{figure*}[t!]
\includegraphics[width=1.65\columnwidth]{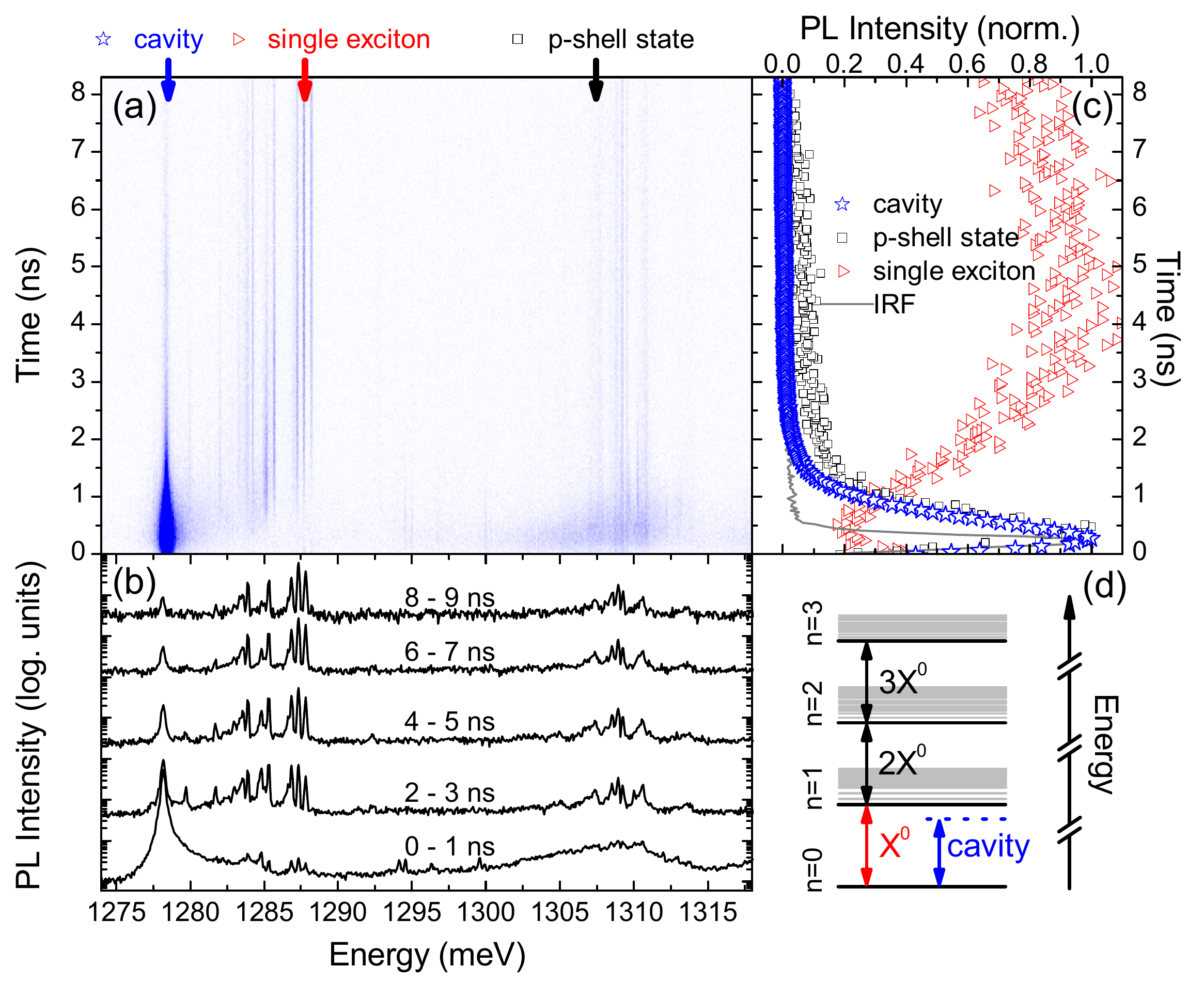}
\caption{\label{figure4} (color online) (a) False color contour plot of the time-resolved photoluminescence intensity of a QD-cavity system as a function of emission wavelength.~\protect\cite{remarkplot} (b) PL spectra at different time delays after the laser pulse, each integrated over $1$~ns. (c) Extracted, normalized photoluminescence intensity from the cavity mode (blue stars), the single exciton (red triangles), and a p-shell transition (black squares) as a function of time delay after the laser pulse. The IRF is plotted as gray solid line. (d) Schematic energy-level diagram (n$\leq 3$) of a neutral quantum dot (drawn after Ref.~\cite{Winger09}).}
\end{figure*}
In order to obtain first ideas about the mechanisms responsible for such strongly non-resonant coupling, we conduct pump power dependent measurements on a QD-cavity system where the cavity mode is at $\Delta E>5$~meV lower energy than the QD s-shell. We plot the corresponding PL spectra in Fig.\ref{figure3}(a) on a logarithmic scale. Here, the incoherent excitation power was increased from $P_{exc}=10$~nW (bottom spectrum) to $P_{exc}=500$~nW (top spectrum). For the lowest pump power, the s-shell of the QD is already clearly visible whilst emission from the p-shell is barely observed. Furthermore, emission from the cavity mode at $E_{cav}=1277.7$~meV is very weak. As we increase the excitation power the emission intensity from the s-shell increases until it saturates for excitation powers in excess of $P\sim100$~nW. At the same time, the emission intensity from the p-shell increases continuously and does not saturate within the plotted power. Emission from the cavity mode increases continuously and, furthermore, does not saturate over the investigated range of excitation powers. The observed power dependence of the cavity mode is, therefore, unrelated to that of the QD s-shell and more closely reflects the QD p-shell.
In Fig.\ref{figure3}(b) we plot the extracted peak intensity of the three selected spectral lines marked in Fig.\ref{figure3}(a) as a function of excitation power on a double-logarithmic scale. These are a single exciton transition (red triangles), a p-shell transition (black squares) and the cavity mode (blue stars). The single exciton transition exhibits a nearly linear power dependence with an exponent of $m_{X}=1.05\pm0.06$, characteristic for a single exciton.~\cite{Finley01,abbarchi09} The selected state from the p-shell exhibits a superlinear power dependence with an exponent of $m_{p-shell}=1.35\pm0.05$, and the cavity mode intensity also increases superlinearly with an exponent of $m_{cav}=1.56\pm0.03$, even faster than the selected p-shell transition. Furthermore, p-shell and cavity mode saturate at a similar excitation power ($\sim1$~$\mu$W), much higher than the single exciton emission ($\sim100$~nW). These observations strongly suggest that the cavity mode emission is related to the multi-exciton emission from the QDs.\cite{Kaniber08b, Winger09, Laucht10b}

%
To obtain further support for the multi-exciton feeding of the cavity mode, time-resolved PL measurements are performed on the same QD-cavity system. We chose the laser power such that the system is excited above saturation of the single exciton level ($P=1000$~nW). Due to the cascaded nature of the population decay in the QD, emission from the single exciton transition is temporally delayed, such that the maximum intensity is not reached until a few nanoseconds after excitation with the laser pulse. At that time the average exciton population in the dot has already decayed close to the single excitation level. In Fig.~\ref{figure4}(a) we plot the time-resolved emission intensity of the QD-cavity system as a function of energy. Emission from the p-shell transitions ($E_{p-shell}=1306-1314$~meV) is observed rapidly after the arrival of the laser excitation pulse at $t=0$~ns and decays within a few nanoseconds.~\cite{lifetime} As discussed above, emission from the s-shell transitions ($E_{s-shell 1}=1283-1288$~meV) is temporally delayed, and exhibits a maximum intensity $\sim4-5$~ns after excitation. Similarly the emission from the cavity mode at $E_{cav}=1278.4$~meV appears rapidly after arrival of the excitation pulse and decays quickly within $\sim1.5$~ns. 
The PL spectra plotted in Fig.~\ref{figure4}(b) show again the time evolution of the QD-cavity system. Here, we integrated the measured time-resolved PL signal over $\Delta t=1$~ns time intervals and present the resulting spectra for the time intervals from $t=0-1$~ns, $t=2-3$~ns, $t=4-5$~ns, $t=6-7$~ns, and $t=8-9$~ns, from bottom to top. During the first nanosecond interval, emission from the QD p-shell and from the cavity mode dominate the spectrum (please note the logarithmic scale). However, the intensity of this emission decreases rapidly and is much weaker, already $2-3$~ns, after the excitation. At later times $t>4$~ns, the dominating peaks of the spectrum are the s-shell transitions, whilst only little signal from the cavity mode is observed. 

In Fig.~\ref{figure4}(c) we plot the integrated, normalized PL intensity of the three states marked with arrows in Fig.~\ref{figure4}(a) as a function of time. Here, we can directly compare the decay transients of the single exciton (red triangles), the p-shell state (black squares), and the cavity mode (blue stars). The solid gray line is the IRF of our experimental setup, measured with the detection tuned to the laser wavelength. It serves as reference for the time when the laser pulse excites the sample and allows us to determine the origin of the time axis. While emission from the cavity mode and the p-shell state occurs immediately after the laser pulse, the emission from the single exciton is delayed and is temporally uncorrelated with the mode emission.

All experimental evidence indicates that the exact mechanism which leads to non-resonant emission of photons into the cavity mode is related to multi-excitons. In Fig.~\ref{figure4}(d) we plot a schematic energy level diagram for a QD-cavity system (redrawn from Ref.~\cite{Winger09}). The black lines, labeled with $n=0$, $n=1$, $n=2$, and $n=3$, indicate the energetically most favorable configurations of $n$ excitons in the QD. The corresponding transitions are labeled with $X^0$, $2X^0$, and $3X^0$. The energy of the cavity mode is assumed to be smaller than that of a single exciton and is schematically represented as a blue, dotted line.
In a more realistic picture, the additional energy levels have to be considered as well. Each of the excited states possesses a number of different possibilities to distribute the charges into different orbital levels and configurations. Each of these different charge configurations has a specific energy that is determined by the competition between the orbital kinetic energy and the Coulomb interactions between confined particles.~\cite{Karrai04} This gives rise to a continuum of QD excited states for each of the multi-excitons. All these states are indicated as gray solid lines in Fig.~\ref{figure4}(d). Transitions between this plethora of states will merge into a quasi-continuum and generate a broad QD background at higher excitation levels.~\cite{Dekel98} Some of the states within this continuum will be in resonance with the cavity mode and their Purcell-enhanced emission can efficiently feed photons into the mode.~\cite{Kaniber08b, Winger09, Laucht10b}

%
%
In summary we have shown PL intensity measurements that demonstrate phonon-mediated feeding of the cavity mode in a strongly coupled QD-cavity system. Cross-correlations measurements reveal an additional feeding mechanism which is active for large detunings $\left|\Delta E\right|>5$~meV. The power dependence of PL intensity measurements as well as time-resolved measurements on the temporal emission of different QD states and the cavity mode show a strong correlation between multi-exciton and cavity mode emission, whilst the emission of single excitons is completely uncorrelated. These results strengthen our understanding of the cavity feeding and, furthermore, may allow us to devise means of exploiting or avoiding it in order to enhance the single photon emission purity of single photon sources or even increase the gain in QD nanocavity lasers.\cite{Strauf06}

We acknowledge financial support of the DFG via the SFB 631, Teilprojekt B3, the German Excellence Initiative via NIM, and the European Union within FP-7 via SOLID.


\begin{thebibliography}{0}%
\makeatletter
\providecommand \@ifxundefined [1]{%
 \@ifx{#1\undefined}
}%
\providecommand \@ifnum [1]{%
 \ifnum #1\expandafter \@firstoftwo
 \else \expandafter \@secondoftwo
 \fi
}%
\providecommand \@ifx [1]{%
 \ifx #1\expandafter \@firstoftwo
 \else \expandafter \@secondoftwo
 \fi
}%
\providecommand \natexlab [1]{#1}%
\providecommand \enquote  [1]{``#1''}%
\providecommand \bibnamefont  [1]{#1}%
\providecommand \bibfnamefont [1]{#1}%
\providecommand \citenamefont [1]{#1}%
\providecommand \href@noop [0]{\@secondoftwo}%
\providecommand \href [0]{\begingroup \@sanitize@url \@href}%
\providecommand \@href[1]{\@@startlink{#1}\@@href}%
\providecommand \@@href[1]{\endgroup#1\@@endlink}%
\providecommand \@sanitize@url [0]{\catcode `\\12\catcode `\$12\catcode
  `\&12\catcode `\#12\catcode `\^12\catcode `\_12\catcode `\%12\relax}%
\providecommand \@@startlink[1]{}%
\providecommand \@@endlink[0]{}%
\providecommand \url  [0]{\begingroup\@sanitize@url \@url }%
\providecommand \@url [1]{\endgroup\@href {#1}{\urlprefix }}%
\providecommand \urlprefix  [0]{URL }%
\providecommand \Eprint [0]{\href }%
\@ifxundefined \urlstyle {%
  \providecommand \doi  [0]{\begingroup \@sanitize@url \@doi}%
  \providecommand \@doi [1]{\endgroup \@@startlink {\doibase
  #1}doi:\discretionary {}{}{}#1\@@endlink }%
}{%
  \providecommand \doi  [0]{doi:\discretionary{}{}{}\begingroup
  \urlstyle{rm}\Url }%
}%
\providecommand \doibase [0]{http://dx.doi.org/}%
\providecommand \Doi [0]{\begingroup \@sanitize@url \@Doi }%
\providecommand \@Doi  [1]{\endgroup\@@startlink{\doibase#1}\@@Doi}%
\providecommand \@@Doi [1]{#1\@@endlink}%
\providecommand \selectlanguage [0]{\@gobble}%
\providecommand \bibinfo  [0]{\@secondoftwo}%
\providecommand \bibfield  [0]{\@secondoftwo}%
\providecommand \translation [1]{[#1]}%
\providecommand \BibitemOpen [0]{}%
\providecommand \bibitemStop [0]{}%
\providecommand \bibitemNoStop [0]{.\EOS\space}%
\providecommand \EOS [0]{\spacefactor3000\relax}%
\providecommand \BibitemShut  [1]{\csname bibitem#1\endcsname}%
\end{thebibliography}%


\begin{thebibliography}{27}%
\makeatletter
\providecommand \@ifxundefined [1]{%
 \@ifx{#1\undefined}
}%
\providecommand \@ifnum [1]{%
 \ifnum #1\expandafter \@firstoftwo
 \else \expandafter \@secondoftwo
 \fi
}%
\providecommand \@ifx [1]{%
 \ifx #1\expandafter \@firstoftwo
 \else \expandafter \@secondoftwo
 \fi
}%
\providecommand \natexlab [1]{#1}%
\providecommand \enquote  [1]{``#1''}%
\providecommand \bibnamefont  [1]{#1}%
\providecommand \bibfnamefont [1]{#1}%
\providecommand \citenamefont [1]{#1}%
\providecommand \href@noop [0]{\@secondoftwo}%
\providecommand \href [0]{\begingroup \@sanitize@url \@href}%
\providecommand \@href[1]{\@@startlink{#1}\@@href}%
\providecommand \@@href[1]{\endgroup#1\@@endlink}%
\providecommand \@sanitize@url [0]{\catcode `\\12\catcode `\$12\catcode
  `\&12\catcode `\#12\catcode `\^12\catcode `\_12\catcode `\%12\relax}%
\providecommand \@@startlink[1]{}%
\providecommand \@@endlink[0]{}%
\providecommand \url  [0]{\begingroup\@sanitize@url \@url }%
\providecommand \@url [1]{\endgroup\@href {#1}{\urlprefix }}%
\providecommand \urlprefix  [0]{URL }%
\providecommand \Eprint [0]{\href }%
\@ifxundefined \urlstyle {%
  \providecommand \doi  [0]{\begingroup \@sanitize@url \@doi}%
  \providecommand \@doi [1]{\endgroup \@@startlink {\doibase
  #1}doi:\discretionary {}{}{}#1\@@endlink }%
}{%
  \providecommand \doi  [0]{doi:\discretionary{}{}{}\begingroup
  \urlstyle{rm}\Url }%
}%
\providecommand \doibase [0]{http://dx.doi.org/}%
\providecommand \Doi [0]{\begingroup \@sanitize@url \@Doi }%
\providecommand \@Doi  [1]{\endgroup\@@startlink{\doibase#1}\@@Doi}%
\providecommand \@@Doi [1]{#1\@@endlink}%
\providecommand \selectlanguage [0]{\@gobble}%
\providecommand \bibinfo  [0]{\@secondoftwo}%
\providecommand \bibfield  [0]{\@secondoftwo}%
\providecommand \translation [1]{[#1]}%
\providecommand \BibitemOpen [0]{}%
\providecommand \bibitemStop [0]{}%
\providecommand \bibitemNoStop [0]{.\EOS\space}%
\providecommand \EOS [0]{\spacefactor3000\relax}%
\providecommand \BibitemShut  [1]{\csname bibitem#1\endcsname}%
\bibitem [{\citenamefont {Vahala}(2003)}]{Vahala03}%
  \BibitemOpen
  \bibfield  {author} {\bibinfo {author} {\bibfnamefont {K.~J.}\ \bibnamefont
  {Vahala}},\ }\bibfield  {title} {\enquote {\bibinfo {title} {Optical
  microcavities},}\ }\href@noop {} {\bibfield  {journal} {\bibinfo  {journal}
  {Nature},\ }\textbf {\bibinfo {volume} {424}},\ \bibinfo {pages} {839--846}
  (\bibinfo {year} {2003})}\BibitemShut {NoStop}%
\bibitem [{\citenamefont {Santori}\ \emph {et~al.}(2002)\citenamefont
  {Santori}, \citenamefont {Fattal}, \citenamefont {Vuckovi\'{c}},
  \citenamefont {Solomon},\ and\ \citenamefont {Yamamoto}}]{Santori02}%
  \BibitemOpen
  \bibfield  {author} {\bibinfo {author} {\bibfnamefont {C.}~\bibnamefont
  {Santori}}, \bibinfo {author} {\bibfnamefont {D.}~\bibnamefont {Fattal}},
  \bibinfo {author} {\bibfnamefont {J.}~\bibnamefont {Vuckovi\'{c}}}, \bibinfo
  {author} {\bibfnamefont {G.~S.}\ \bibnamefont {Solomon}}, \ and\ \bibinfo
  {author} {\bibfnamefont {Y.}~\bibnamefont {Yamamoto}},\ }\bibfield  {title}
  {\enquote {\bibinfo {title} {Indistinguishable photons from a single-photon
  device},}\ }\href@noop {} {\bibfield  {journal} {\bibinfo  {journal}
  {Nature},\ }\textbf {\bibinfo {volume} {419}},\ \bibinfo {pages} {594}
  (\bibinfo {year} {2002})}\BibitemShut {NoStop}%
\bibitem [{\citenamefont {Englund}\ \emph {et~al.}(2007)\citenamefont
  {Englund}, \citenamefont {Faraon}, \citenamefont {Fushman}, \citenamefont
  {Stoltz}, \citenamefont {Petroff},\ and\ \citenamefont
  {Vuckovic}}]{Englund07}%
  \BibitemOpen
  \bibfield  {author} {\bibinfo {author} {\bibfnamefont {D.}~\bibnamefont
  {Englund}}, \bibinfo {author} {\bibfnamefont {A.}~\bibnamefont {Faraon}},
  \bibinfo {author} {\bibfnamefont {I.}~\bibnamefont {Fushman}}, \bibinfo
  {author} {\bibfnamefont {N.}~\bibnamefont {Stoltz}}, \bibinfo {author}
  {\bibfnamefont {P.}~\bibnamefont {Petroff}}, \ and\ \bibinfo {author}
  {\bibfnamefont {J.}~\bibnamefont {Vuckovic}},\ }\bibfield  {title} {\enquote
  {\bibinfo {title} {Controlling cavity reflectivity with a single quantum
  dot},}\ }\href@noop {} {\bibfield  {journal} {\bibinfo  {journal} {Nature},\
  }\textbf {\bibinfo {volume} {450}},\ \bibinfo {pages} {857--861} (\bibinfo
  {year} {2007})}\BibitemShut {NoStop}%
\bibitem [{\citenamefont {Strauf}\ \emph {et~al.}(2006)\citenamefont {Strauf},
  \citenamefont {Hennessy}, \citenamefont {Rakher}, \citenamefont {Choi},
  \citenamefont {Badolato}, \citenamefont {Andreani}, \citenamefont {Hu},
  \citenamefont {Petroff},\ and\ \citenamefont {Bouwmeester}}]{Strauf06}%
  \BibitemOpen
  \bibfield  {author} {\bibinfo {author} {\bibfnamefont {S.}~\bibnamefont
  {Strauf}}, \bibinfo {author} {\bibfnamefont {K.}~\bibnamefont {Hennessy}},
  \bibinfo {author} {\bibfnamefont {M.~T.}\ \bibnamefont {Rakher}}, \bibinfo
  {author} {\bibfnamefont {Y.~S.}\ \bibnamefont {Choi}}, \bibinfo {author}
  {\bibfnamefont {A.}~\bibnamefont {Badolato}}, \bibinfo {author}
  {\bibfnamefont {L.~C.}\ \bibnamefont {Andreani}}, \bibinfo {author}
  {\bibfnamefont {E.~L.}\ \bibnamefont {Hu}}, \bibinfo {author} {\bibfnamefont
  {P.~M.}\ \bibnamefont {Petroff}}, \ and\ \bibinfo {author} {\bibfnamefont
  {D.}~\bibnamefont {Bouwmeester}},\ }\bibfield  {title} {\enquote {\bibinfo
  {title} {Self-tuned quantum dot gain in photonic crystal lasers},}\
  }\href@noop {} {\bibfield  {journal} {\bibinfo  {journal} {Phys. Rev.
  Lett.},\ }\textbf {\bibinfo {volume} {96}},\ \bibinfo {pages} {127404}
  (\bibinfo {year} {2006})}\BibitemShut {NoStop}%
\bibitem [{\citenamefont {Hohenester}\ \emph {et~al.}(2009)\citenamefont
  {Hohenester}, \citenamefont {Laucht}, \citenamefont {Kaniber}, \citenamefont
  {Hauke}, \citenamefont {Neumann}, \citenamefont {Mohtashami}, \citenamefont
  {Seliger}, \citenamefont {Bichler},\ and\ \citenamefont
  {Finley}}]{Hohenester09}%
  \BibitemOpen
  \bibfield  {author} {\bibinfo {author} {\bibfnamefont {U.}~\bibnamefont
  {Hohenester}}, \bibinfo {author} {\bibfnamefont {A.}~\bibnamefont {Laucht}},
  \bibinfo {author} {\bibfnamefont {M.}~\bibnamefont {Kaniber}}, \bibinfo
  {author} {\bibfnamefont {N.}~\bibnamefont {Hauke}}, \bibinfo {author}
  {\bibfnamefont {A.}~\bibnamefont {Neumann}}, \bibinfo {author} {\bibfnamefont
  {A.}~\bibnamefont {Mohtashami}}, \bibinfo {author} {\bibfnamefont
  {M.}~\bibnamefont {Seliger}}, \bibinfo {author} {\bibfnamefont
  {M.}~\bibnamefont {Bichler}}, \ and\ \bibinfo {author} {\bibfnamefont
  {J.~J.}\ \bibnamefont {Finley}},\ }\bibfield  {title} {\enquote {\bibinfo
  {title} {Phonon-assisted transitions from quantum dot excitons to cavity
  photons},}\ }\href@noop {} {\bibfield  {journal} {\bibinfo  {journal} {Phys.
  Rev. B},\ }\textbf {\bibinfo {volume} {80}},\ \bibinfo {pages} {201311(R)}
  (\bibinfo {year} {2009})}\BibitemShut {NoStop}%
\bibitem [{\citenamefont {Suffczy\'nski}\ \emph {et~al.}(2009)\citenamefont
  {Suffczy\'nski}, \citenamefont {Dousse}, \citenamefont {Lema\^itre},
  \citenamefont {Sagnes}, \citenamefont {Lanco}, \citenamefont {Bloch},
  \citenamefont {Voisin},\ and\ \citenamefont {Senellart}}]{Suffcynski09}%
  \BibitemOpen
  \bibfield  {author} {\bibinfo {author} {\bibfnamefont {J.}~\bibnamefont
  {Suffczy\'nski}}, \bibinfo {author} {\bibfnamefont {A.}~\bibnamefont
  {Dousse}}, \bibinfo {author} {\bibfnamefont {A.}~\bibnamefont {Lema\^itre}},
  \bibinfo {author} {\bibfnamefont {I.}~\bibnamefont {Sagnes}}, \bibinfo
  {author} {\bibfnamefont {L.}~\bibnamefont {Lanco}}, \bibinfo {author}
  {\bibfnamefont {J.}~\bibnamefont {Bloch}}, \bibinfo {author} {\bibfnamefont
  {P.}~\bibnamefont {Voisin}}, \ and\ \bibinfo {author} {\bibfnamefont
  {P.}~\bibnamefont {Senellart}},\ }\bibfield  {title} {\enquote {\bibinfo
  {title} {Origin of the optical emission within the cavity mode of coupled
  quantum-dot-cavity systems},}\ }\href@noop {} {\bibfield  {journal} {\bibinfo
   {journal} {Phys. Rev. Lett.},\ }\textbf {\bibinfo {volume} {103}},\ \bibinfo
  {pages} {027401} (\bibinfo {year} {2009})}\BibitemShut {NoStop}%
\bibitem [{\citenamefont {Hohenester}(2010)}]{Hohenester10}%
  \BibitemOpen
  \bibfield  {author} {\bibinfo {author} {\bibfnamefont {U.}~\bibnamefont
  {Hohenester}},\ }\bibfield  {title} {\enquote {\bibinfo {title} {Cavity
  quantum electrodynamics with semiconductor quantum dots: Role of
  phonon-assisted cavity feeding},}\ }\href@noop {} {\bibfield  {journal}
  {\bibinfo  {journal} {Phys. Rev. B},\ }\textbf {\bibinfo {volume} {81}},\
  \bibinfo {pages} {155303} (\bibinfo {year} {2010})}\BibitemShut {NoStop}%
\bibitem [{\citenamefont {Ota}\ \emph {et~al.}(2009)\citenamefont {Ota},
  \citenamefont {Iwamoto}, \citenamefont {Kumagai},\ and\ \citenamefont
  {Arakawa}}]{Ota09}%
  \BibitemOpen
  \bibfield  {author} {\bibinfo {author} {\bibfnamefont {Y.}~\bibnamefont
  {Ota}}, \bibinfo {author} {\bibfnamefont {S.}~\bibnamefont {Iwamoto}},
  \bibinfo {author} {\bibfnamefont {N.}~\bibnamefont {Kumagai}}, \ and\
  \bibinfo {author} {\bibfnamefont {Y.}~\bibnamefont {Arakawa}},\ }\bibfield
  {title} {\enquote {\bibinfo {title} {Impact of electron-phonon interactions
  on quantum-dot cavity quantum electrodynamics},}\ }\href@noop {} {\bibfield
  {journal} {\bibinfo  {journal} {arXiv.org:0908.0788}} (\bibinfo {year}
  {2009})}\BibitemShut {NoStop}%
\bibitem [{\citenamefont {Hennessy}\ \emph {et~al.}(2007)\citenamefont
  {Hennessy}, \citenamefont {Badolato}, \citenamefont {Winger}, \citenamefont
  {Gerace}, \citenamefont {Atature}, \citenamefont {Gulde}, \citenamefont
  {Falt}, \citenamefont {Hu},\ and\ \citenamefont {Imamo\u{g}lu}}]{Hennessy07}%
  \BibitemOpen
  \bibfield  {author} {\bibinfo {author} {\bibfnamefont {K.}~\bibnamefont
  {Hennessy}}, \bibinfo {author} {\bibfnamefont {A.}~\bibnamefont {Badolato}},
  \bibinfo {author} {\bibfnamefont {M.}~\bibnamefont {Winger}}, \bibinfo
  {author} {\bibfnamefont {D.}~\bibnamefont {Gerace}}, \bibinfo {author}
  {\bibfnamefont {M.}~\bibnamefont {Atature}}, \bibinfo {author} {\bibfnamefont
  {S.}~\bibnamefont {Gulde}}, \bibinfo {author} {\bibfnamefont
  {S.}~\bibnamefont {Falt}}, \bibinfo {author} {\bibfnamefont {E.~L.}\
  \bibnamefont {Hu}}, \ and\ \bibinfo {author} {\bibfnamefont {A.}~\bibnamefont
  {Imamo\u{g}lu}},\ }\bibfield  {title} {\enquote {\bibinfo {title} {Quantum
  nature of a strongly coupled single quantum dot-cavity system},}\ }\href@noop
  {} {\bibfield  {journal} {\bibinfo  {journal} {Nature},\ }\textbf {\bibinfo
  {volume} {445}},\ \bibinfo {pages} {896--899} (\bibinfo {year}
  {2007})}\BibitemShut {NoStop}%
\bibitem [{\citenamefont {Press}\ \emph {et~al.}(2007)\citenamefont {Press},
  \citenamefont {Gotzinger}, \citenamefont {Reitzenstein}, \citenamefont
  {Hofmann}, \citenamefont {Loffler}, \citenamefont {Kamp}, \citenamefont
  {Forchel},\ and\ \citenamefont {Yamamoto}}]{Press07}%
  \BibitemOpen
  \bibfield  {author} {\bibinfo {author} {\bibfnamefont {D.}~\bibnamefont
  {Press}}, \bibinfo {author} {\bibfnamefont {S.}~\bibnamefont {Gotzinger}},
  \bibinfo {author} {\bibfnamefont {S.}~\bibnamefont {Reitzenstein}}, \bibinfo
  {author} {\bibfnamefont {C.}~\bibnamefont {Hofmann}}, \bibinfo {author}
  {\bibfnamefont {A.}~\bibnamefont {Loffler}}, \bibinfo {author} {\bibfnamefont
  {M.}~\bibnamefont {Kamp}}, \bibinfo {author} {\bibfnamefont {A.}~\bibnamefont
  {Forchel}}, \ and\ \bibinfo {author} {\bibfnamefont {Y.}~\bibnamefont
  {Yamamoto}},\ }\bibfield  {title} {\enquote {\bibinfo {title} {Photon
  antibunching from a single quantum-dot-microcavity system in the strong
  coupling regime},}\ }\href@noop {} {\bibfield  {journal} {\bibinfo  {journal}
  {Phys. Rev. Lett.},\ }\textbf {\bibinfo {volume} {98}},\ \bibinfo {pages}
  {117402} (\bibinfo {year} {2007})}\BibitemShut {NoStop}%
\bibitem [{\citenamefont {Kaniber}\ \emph
  {et~al.}(2008){\natexlab{a}}\citenamefont {Kaniber}, \citenamefont {Laucht},
  \citenamefont {Neumann}, \citenamefont {Villas-Boas}, \citenamefont
  {Bichler}, \citenamefont {Amann},\ and\ \citenamefont {Finley}}]{Kaniber08b}%
  \BibitemOpen
  \bibfield  {author} {\bibinfo {author} {\bibfnamefont {M.}~\bibnamefont
  {Kaniber}}, \bibinfo {author} {\bibfnamefont {A.}~\bibnamefont {Laucht}},
  \bibinfo {author} {\bibfnamefont {A.}~\bibnamefont {Neumann}}, \bibinfo
  {author} {\bibfnamefont {J.~M.}\ \bibnamefont {Villas-Boas}}, \bibinfo
  {author} {\bibfnamefont {M.}~\bibnamefont {Bichler}}, \bibinfo {author}
  {\bibfnamefont {M.-C.}\ \bibnamefont {Amann}}, \ and\ \bibinfo {author}
  {\bibfnamefont {J.~J.}\ \bibnamefont {Finley}},\ }\bibfield  {title}
  {\enquote {\bibinfo {title} {Investigation of the nonresonant dot-cavity
  coupling in two-dimensional photonic crystal nanocavities},}\ }\href@noop {}
  {\bibfield  {journal} {\bibinfo  {journal} {Phys. Rev. B},\ }\textbf
  {\bibinfo {volume} {77}},\ \bibinfo {pages} {161303(R)} (\bibinfo {year}
  {2008}{\natexlab{a}})}\BibitemShut {NoStop}%
\bibitem [{\citenamefont {Winger}\ \emph {et~al.}(2009)\citenamefont {Winger},
  \citenamefont {Volz}, \citenamefont {Tarel}, \citenamefont {Portolan},
  \citenamefont {Badolato}, \citenamefont {Hennessy}, \citenamefont {Hu},
  \citenamefont {Beveratos}, \citenamefont {Finley}, \citenamefont {Savona},\
  and\ \citenamefont {Imamo\u{g}lu}}]{Winger09}%
  \BibitemOpen
  \bibfield  {author} {\bibinfo {author} {\bibfnamefont {M.}~\bibnamefont
  {Winger}}, \bibinfo {author} {\bibfnamefont {T.}~\bibnamefont {Volz}},
  \bibinfo {author} {\bibfnamefont {G.}~\bibnamefont {Tarel}}, \bibinfo
  {author} {\bibfnamefont {S.}~\bibnamefont {Portolan}}, \bibinfo {author}
  {\bibfnamefont {A.}~\bibnamefont {Badolato}}, \bibinfo {author}
  {\bibfnamefont {K.~J.}\ \bibnamefont {Hennessy}}, \bibinfo {author}
  {\bibfnamefont {E.~L.}\ \bibnamefont {Hu}}, \bibinfo {author} {\bibfnamefont
  {A.}~\bibnamefont {Beveratos}}, \bibinfo {author} {\bibfnamefont
  {J.}~\bibnamefont {Finley}}, \bibinfo {author} {\bibfnamefont
  {V.}~\bibnamefont {Savona}}, \ and\ \bibinfo {author} {\bibfnamefont
  {A.}~\bibnamefont {Imamo\u{g}lu}},\ }\bibfield  {title} {\enquote {\bibinfo
  {title} {Explanation of photon correlations in the far-off-resonance optical
  emission from a quantum-dot--cavity system},}\ }\href@noop {} {\bibfield
  {journal} {\bibinfo  {journal} {Phys. Rev. Lett.},\ }\textbf {\bibinfo
  {volume} {103}},\ \bibinfo {pages} {207403} (\bibinfo {year}
  {2009})}\BibitemShut {NoStop}%
\bibitem [{\citenamefont {{Laucht}}\ \emph {et~al.}(2010)\citenamefont
  {{Laucht}}, \citenamefont {{Kaniber}}, \citenamefont {{Mohtashami}},
  \citenamefont {{Hauke}}, \citenamefont {{Bichler}},\ and\ \citenamefont
  {{Finley}}}]{Laucht10b}%
  \BibitemOpen
  \bibfield  {author} {\bibinfo {author} {\bibfnamefont {A.}~\bibnamefont
  {{Laucht}}}, \bibinfo {author} {\bibfnamefont {M.}~\bibnamefont {{Kaniber}}},
  \bibinfo {author} {\bibfnamefont {A.}~\bibnamefont {{Mohtashami}}}, \bibinfo
  {author} {\bibfnamefont {N.}~\bibnamefont {{Hauke}}}, \bibinfo {author}
  {\bibfnamefont {M.}~\bibnamefont {{Bichler}}}, \ and\ \bibinfo {author}
  {\bibfnamefont {J.~J.}\ \bibnamefont {{Finley}}},\ }\bibfield  {title}
  {\enquote {\bibinfo {title} {{Temporal Monitoring of Non-resonant Feeding of
  Semiconductor Nanocavity Modes by Quantum Dot Multiexciton Transitions}},}\
  }\href@noop {} {\bibfield  {journal} {\bibinfo  {journal} {Phys. Rev. B},\
  }\textbf {\bibinfo {volume} {81}},\ \bibinfo {pages} {241302(R)} (\bibinfo
  {year} {2010})}\BibitemShut {NoStop}%
\bibitem [{\citenamefont {Laucht}\ \emph
  {et~al.}(2009){\natexlab{a}}\citenamefont {Laucht}, \citenamefont {Hofbauer},
  \citenamefont {Hauke}, \citenamefont {Angele}, \citenamefont {Stobbe},
  \citenamefont {Kaniber}, \citenamefont {B\"ohm}, \citenamefont {Lodahl},
  \citenamefont {Amann},\ and\ \citenamefont {Finley}}]{Laucht09}%
  \BibitemOpen
  \bibfield  {author} {\bibinfo {author} {\bibfnamefont {A.}~\bibnamefont
  {Laucht}}, \bibinfo {author} {\bibfnamefont {F.}~\bibnamefont {Hofbauer}},
  \bibinfo {author} {\bibfnamefont {N.}~\bibnamefont {Hauke}}, \bibinfo
  {author} {\bibfnamefont {J.}~\bibnamefont {Angele}}, \bibinfo {author}
  {\bibfnamefont {S.}~\bibnamefont {Stobbe}}, \bibinfo {author} {\bibfnamefont
  {M.}~\bibnamefont {Kaniber}}, \bibinfo {author} {\bibfnamefont
  {G.}~\bibnamefont {B\"ohm}}, \bibinfo {author} {\bibfnamefont
  {P.}~\bibnamefont {Lodahl}}, \bibinfo {author} {\bibfnamefont {M.-C.}\
  \bibnamefont {Amann}}, \ and\ \bibinfo {author} {\bibfnamefont {J.~J.}\
  \bibnamefont {Finley}},\ }\bibfield  {title} {\enquote {\bibinfo {title}
  {Electrical control of spontaneous emission and strong coupling for a single
  quantum dot},}\ }\href@noop {} {\bibfield  {journal} {\bibinfo  {journal}
  {New J. Phys.},\ }\textbf {\bibinfo {volume} {11}},\ \bibinfo {pages}
  {023034} (\bibinfo {year} {2009}{\natexlab{a}})}\BibitemShut {NoStop}%
\bibitem [{\citenamefont {Laucht}\ \emph
  {et~al.}(2009){\natexlab{b}}\citenamefont {Laucht}, \citenamefont {Hauke},
  \citenamefont {Villas-B\^{o}as}, \citenamefont {Hofbauer}, \citenamefont
  {B\"{o}hm}, \citenamefont {Kaniber},\ and\ \citenamefont
  {Finley}}]{Laucht09b}%
  \BibitemOpen
  \bibfield  {author} {\bibinfo {author} {\bibfnamefont {A.}~\bibnamefont
  {Laucht}}, \bibinfo {author} {\bibfnamefont {N.}~\bibnamefont {Hauke}},
  \bibinfo {author} {\bibfnamefont {J.~M.}\ \bibnamefont {Villas-B\^{o}as}},
  \bibinfo {author} {\bibfnamefont {F.}~\bibnamefont {Hofbauer}}, \bibinfo
  {author} {\bibfnamefont {G.}~\bibnamefont {B\"{o}hm}}, \bibinfo {author}
  {\bibfnamefont {M.}~\bibnamefont {Kaniber}}, \ and\ \bibinfo {author}
  {\bibfnamefont {J.~J.}\ \bibnamefont {Finley}},\ }\bibfield  {title}
  {\enquote {\bibinfo {title} {Dephasing of exciton polaritons in photoexcited
  ingaas quantum dots in gaas nanocavities},}\ }\href@noop {} {\bibfield
  {journal} {\bibinfo  {journal} {Phys. Rev. Lett.},\ }\textbf {\bibinfo
  {volume} {103}},\ \bibinfo {eid} {087405} (\bibinfo {year}
  {2009}{\natexlab{b}})}\BibitemShut {NoStop}%
\bibitem [{\citenamefont {Chauvin}\ \emph {et~al.}(2009)\citenamefont
  {Chauvin}, \citenamefont {Zinoni}, \citenamefont {Francardi}, \citenamefont
  {Gerardino}, \citenamefont {Balet}, \citenamefont {Alloing}, \citenamefont
  {Li},\ and\ \citenamefont {Fiore}}]{Chauvin09}%
  \BibitemOpen
  \bibfield  {author} {\bibinfo {author} {\bibfnamefont {N.}~\bibnamefont
  {Chauvin}}, \bibinfo {author} {\bibfnamefont {C.}~\bibnamefont {Zinoni}},
  \bibinfo {author} {\bibfnamefont {M.}~\bibnamefont {Francardi}}, \bibinfo
  {author} {\bibfnamefont {A.}~\bibnamefont {Gerardino}}, \bibinfo {author}
  {\bibfnamefont {L.}~\bibnamefont {Balet}}, \bibinfo {author} {\bibfnamefont
  {B.}~\bibnamefont {Alloing}}, \bibinfo {author} {\bibfnamefont
  {L.}~\bibnamefont {Li}}, \ and\ \bibinfo {author} {\bibfnamefont
  {A.}~\bibnamefont {Fiore}},\ }\bibfield  {title} {\enquote {\bibinfo {title}
  {Controlling the charge environment of single quantum dots in a
  photonic-crystal cavity},}\ }\href@noop {} {\bibfield  {journal} {\bibinfo
  {journal} {Phys. Rev. B},\ }\textbf {\bibinfo {volume} {80}},\ \bibinfo
  {pages} {241306(R)} (\bibinfo {year} {2009})}\BibitemShut {NoStop}%
\bibitem [{\citenamefont {Hofbauer}\ \emph {et~al.}(2007)\citenamefont
  {Hofbauer}, \citenamefont {Grimminger}, \citenamefont {Angele}, \citenamefont
  {B\"ohm}, \citenamefont {Meyer}, \citenamefont {Amann},\ and\ \citenamefont
  {Finley}}]{Hofbauer07}%
  \BibitemOpen
  \bibfield  {author} {\bibinfo {author} {\bibfnamefont {F.}~\bibnamefont
  {Hofbauer}}, \bibinfo {author} {\bibfnamefont {S.}~\bibnamefont
  {Grimminger}}, \bibinfo {author} {\bibfnamefont {J.}~\bibnamefont {Angele}},
  \bibinfo {author} {\bibfnamefont {G.}~\bibnamefont {B\"ohm}}, \bibinfo
  {author} {\bibfnamefont {R.}~\bibnamefont {Meyer}}, \bibinfo {author}
  {\bibfnamefont {M.~C.}\ \bibnamefont {Amann}}, \ and\ \bibinfo {author}
  {\bibfnamefont {J.~J.}\ \bibnamefont {Finley}},\ }\bibfield  {title}
  {\enquote {\bibinfo {title} {Electrically probing photonic bandgap phenomena
  in contacted defect nanocavities},}\ }\href@noop {} {\bibfield  {journal}
  {\bibinfo  {journal} {Appl. Phys. Lett.},\ }\textbf {\bibinfo {volume}
  {91}},\ \bibinfo {pages} {201111} (\bibinfo {year} {2007})}\BibitemShut
  {NoStop}%
\bibitem [{\citenamefont {Akahane}\ \emph {et~al.}(2003)\citenamefont
  {Akahane}, \citenamefont {Asano}, \citenamefont {Song},\ and\ \citenamefont
  {Noda}}]{Akahane03}%
  \BibitemOpen
  \bibfield  {author} {\bibinfo {author} {\bibfnamefont {Y.}~\bibnamefont
  {Akahane}}, \bibinfo {author} {\bibfnamefont {T.}~\bibnamefont {Asano}},
  \bibinfo {author} {\bibfnamefont {B.~S.}\ \bibnamefont {Song}}, \ and\
  \bibinfo {author} {\bibfnamefont {S.}~\bibnamefont {Noda}},\ }\bibfield
  {title} {\enquote {\bibinfo {title} {High-q photonic nanocavity in a
  two-dimensional photonic crystal},}\ }\href@noop {} {\bibfield  {journal}
  {\bibinfo  {journal} {Nature},\ }\textbf {\bibinfo {volume} {425}},\ \bibinfo
  {pages} {944--947} (\bibinfo {year} {2003})}\BibitemShut {NoStop}%
\bibitem [{\citenamefont {Kaniber}\ \emph {et~al.}(2009)\citenamefont
  {Kaniber}, \citenamefont {Neumann}, \citenamefont {Laucht}, \citenamefont
  {Bichler}, \citenamefont {Amann},\ and\ \citenamefont {Finley}}]{Kaniber09}%
  \BibitemOpen
  \bibfield  {author} {\bibinfo {author} {\bibfnamefont {M.}~\bibnamefont
  {Kaniber}}, \bibinfo {author} {\bibfnamefont {A.}~\bibnamefont {Neumann}},
  \bibinfo {author} {\bibfnamefont {A.}~\bibnamefont {Laucht}}, \bibinfo
  {author} {\bibfnamefont {M.}~\bibnamefont {Bichler}}, \bibinfo {author}
  {\bibfnamefont {M.-C.}\ \bibnamefont {Amann}}, \ and\ \bibinfo {author}
  {\bibfnamefont {J.~J.}\ \bibnamefont {Finley}},\ }\bibfield  {title}
  {\enquote {\bibinfo {title} {Cavity-resonant excitation for efficient single
  photon generation},}\ }\href@noop {} {\bibfield  {journal} {\bibinfo
  {journal} {New J. Phys.},\ }\textbf {\bibinfo {volume} {11}},\ \bibinfo
  {pages} {013031} (\bibinfo {year} {2009})}\BibitemShut {NoStop}%
\bibitem [{rem()}]{remarkplot}%
  \BibitemOpen
  \href@noop {} {}\bibinfo {note} {The spectrally broad mode emission at early
  times is an artifact of plotting and not a physical effect. Here, the linear
  color scale was chosen such that features of low intensity are well visible
  and the intensity above a certain threshold is plotted with the same
  color.}\BibitemShut {Stop}%
\bibitem [{\citenamefont {Finley}\ \emph {et~al.}(2001)\citenamefont {Finley},
  \citenamefont {Ashmore}, \citenamefont {Lemaitre}, \citenamefont {Mowbray},
  \citenamefont {Skolnick}, \citenamefont {Itskevich}, \citenamefont {Maksym},
  \citenamefont {Hopkinson},\ and\ \citenamefont {Krauss}}]{Finley01}%
  \BibitemOpen
  \bibfield  {author} {\bibinfo {author} {\bibfnamefont {J.~J.}\ \bibnamefont
  {Finley}}, \bibinfo {author} {\bibfnamefont {A.~D.}\ \bibnamefont {Ashmore}},
  \bibinfo {author} {\bibfnamefont {A.}~\bibnamefont {Lemaitre}}, \bibinfo
  {author} {\bibfnamefont {D.~J.}\ \bibnamefont {Mowbray}}, \bibinfo {author}
  {\bibfnamefont {M.~S.}\ \bibnamefont {Skolnick}}, \bibinfo {author}
  {\bibfnamefont {I.~E.}\ \bibnamefont {Itskevich}}, \bibinfo {author}
  {\bibfnamefont {P.~A.}\ \bibnamefont {Maksym}}, \bibinfo {author}
  {\bibfnamefont {M.}~\bibnamefont {Hopkinson}}, \ and\ \bibinfo {author}
  {\bibfnamefont {T.~F.}\ \bibnamefont {Krauss}},\ }\bibfield  {title}
  {\enquote {\bibinfo {title} {Charged and neutral exciton complexes in
  individual self-assembled in(ga)as quantum dots},}\ }\href@noop {} {\bibfield
   {journal} {\bibinfo  {journal} {Phys. Rev. B},\ }\textbf {\bibinfo {volume}
  {63}},\ \bibinfo {pages} {161305(R)} (\bibinfo {year} {2001})}\BibitemShut
  {NoStop}%
\bibitem [{\citenamefont {Abbarchi}\ \emph {et~al.}(2009)\citenamefont
  {Abbarchi}, \citenamefont {Mastrandrea}, \citenamefont {Kuroda},
  \citenamefont {Mano}, \citenamefont {Vinattieri}, \citenamefont {Sakoda},\
  and\ \citenamefont {Gurioli}}]{abbarchi09}%
  \BibitemOpen
  \bibfield  {author} {\bibinfo {author} {\bibfnamefont {M.}~\bibnamefont
  {Abbarchi}}, \bibinfo {author} {\bibfnamefont {C.}~\bibnamefont
  {Mastrandrea}}, \bibinfo {author} {\bibfnamefont {T.}~\bibnamefont {Kuroda}},
  \bibinfo {author} {\bibfnamefont {T.}~\bibnamefont {Mano}}, \bibinfo {author}
  {\bibfnamefont {A.}~\bibnamefont {Vinattieri}}, \bibinfo {author}
  {\bibfnamefont {K.}~\bibnamefont {Sakoda}}, \ and\ \bibinfo {author}
  {\bibfnamefont {M.}~\bibnamefont {Gurioli}},\ }\bibfield  {title} {\enquote
  {\bibinfo {title} {Poissonian statistics of excitonic complexes in quantum
  dots},}\ }\href@noop {} {\bibfield  {journal} {\bibinfo  {journal} {J. Appl.
  Phys.},\ }\textbf {\bibinfo {volume} {106}},\ \bibinfo {pages} {053504}
  (\bibinfo {year} {2009})}\BibitemShut {NoStop}%
\bibitem [{lif()}]{lifetime}%
  \BibitemOpen
  \href@noop {} {}\bibinfo {note} {Here, the lifetime of all transitions is
  lengthened due to the two dimensional photonic
  bandgap~\cite{Kaniber07,Kaniber08a}.}\BibitemShut {Stop}%
\bibitem [{\citenamefont {Karrai}\ \emph {et~al.}(2004)\citenamefont {Karrai},
  \citenamefont {Warburton}, \citenamefont {Schulhauser}, \citenamefont
  {H\"ogele}, \citenamefont {Urbaszek}, \citenamefont {McGhee}, \citenamefont
  {Govorov}, \citenamefont {Garcia}, \citenamefont {Gerardot},\ and\
  \citenamefont {Petroff}}]{Karrai04}%
  \BibitemOpen
  \bibfield  {author} {\bibinfo {author} {\bibfnamefont {K.}~\bibnamefont
  {Karrai}}, \bibinfo {author} {\bibfnamefont {R.}~\bibnamefont {Warburton}},
  \bibinfo {author} {\bibfnamefont {C.}~\bibnamefont {Schulhauser}}, \bibinfo
  {author} {\bibfnamefont {A.}~\bibnamefont {H\"ogele}}, \bibinfo {author}
  {\bibfnamefont {B.}~\bibnamefont {Urbaszek}}, \bibinfo {author}
  {\bibfnamefont {E.}~\bibnamefont {McGhee}}, \bibinfo {author} {\bibfnamefont
  {A.}~\bibnamefont {Govorov}}, \bibinfo {author} {\bibfnamefont
  {J.}~\bibnamefont {Garcia}}, \bibinfo {author} {\bibfnamefont
  {B.}~\bibnamefont {Gerardot}}, \ and\ \bibinfo {author} {\bibfnamefont
  {P.}~\bibnamefont {Petroff}},\ }\bibfield  {title} {\enquote {\bibinfo
  {title} {{Hybridization of electronic states in quantum dots through photon
  emission}},}\ }\href@noop {} {\bibfield  {journal} {\bibinfo  {journal}
  {{Nature}},\ }\textbf {\bibinfo {volume} {{427}}},\ \bibinfo {pages} {{135}}
  (\bibinfo {year} {{2004}})}\BibitemShut {NoStop}%
\bibitem [{\citenamefont {Dekel}\ \emph {et~al.}(1998)\citenamefont {Dekel},
  \citenamefont {Gershoni}, \citenamefont {Ehrenfreund}, \citenamefont
  {Spektor}, \citenamefont {Garcia},\ and\ \citenamefont {Petroff}}]{Dekel98}%
  \BibitemOpen
  \bibfield  {author} {\bibinfo {author} {\bibfnamefont {E.}~\bibnamefont
  {Dekel}}, \bibinfo {author} {\bibfnamefont {D.}~\bibnamefont {Gershoni}},
  \bibinfo {author} {\bibfnamefont {E.}~\bibnamefont {Ehrenfreund}}, \bibinfo
  {author} {\bibfnamefont {D.}~\bibnamefont {Spektor}}, \bibinfo {author}
  {\bibfnamefont {J.~M.}\ \bibnamefont {Garcia}}, \ and\ \bibinfo {author}
  {\bibfnamefont {P.~M.}\ \bibnamefont {Petroff}},\ }\bibfield  {title}
  {\enquote {\bibinfo {title} {Multiexciton spectroscopy of a single
  self-assembled quantum dot},}\ }\href@noop {} {\bibfield  {journal} {\bibinfo
   {journal} {Phys. Rev. Lett.},\ }\textbf {\bibinfo {volume} {80}},\ \bibinfo
  {pages} {4991--4994} (\bibinfo {year} {1998})}\BibitemShut {NoStop}%
\bibitem [{\citenamefont {Kaniber}\ \emph {et~al.}(2007)\citenamefont
  {Kaniber}, \citenamefont {Kress}, \citenamefont {Laucht}, \citenamefont
  {Bichler}, \citenamefont {Meyer}, \citenamefont {Amann},\ and\ \citenamefont
  {Finley}}]{Kaniber07}%
  \BibitemOpen
  \bibfield  {author} {\bibinfo {author} {\bibfnamefont {M.}~\bibnamefont
  {Kaniber}}, \bibinfo {author} {\bibfnamefont {A.}~\bibnamefont {Kress}},
  \bibinfo {author} {\bibfnamefont {A.}~\bibnamefont {Laucht}}, \bibinfo
  {author} {\bibfnamefont {M.}~\bibnamefont {Bichler}}, \bibinfo {author}
  {\bibfnamefont {R.}~\bibnamefont {Meyer}}, \bibinfo {author} {\bibfnamefont
  {M.~C.}\ \bibnamefont {Amann}}, \ and\ \bibinfo {author} {\bibfnamefont
  {J.~J.}\ \bibnamefont {Finley}},\ }\bibfield  {title} {\enquote {\bibinfo
  {title} {Efficient spatial redistribution of quantum dot spontaneous emission
  from two-dimensional photonic crystals},}\ }\href@noop {} {\bibfield
  {journal} {\bibinfo  {journal} {Appl. Phys. Lett.},\ }\textbf {\bibinfo
  {volume} {91}},\ \bibinfo {pages} {061106} (\bibinfo {year}
  {2007})}\BibitemShut {NoStop}%
\bibitem [{\citenamefont {Kaniber}\ \emph
  {et~al.}(2008){\natexlab{b}}\citenamefont {Kaniber}, \citenamefont {Laucht},
  \citenamefont {H\"{u}rlimann}, \citenamefont {Bichler}, \citenamefont
  {Meyer}, \citenamefont {Amann},\ and\ \citenamefont {Finley}}]{Kaniber08a}%
  \BibitemOpen
  \bibfield  {author} {\bibinfo {author} {\bibfnamefont {M.}~\bibnamefont
  {Kaniber}}, \bibinfo {author} {\bibfnamefont {A.}~\bibnamefont {Laucht}},
  \bibinfo {author} {\bibfnamefont {T.}~\bibnamefont {H\"{u}rlimann}}, \bibinfo
  {author} {\bibfnamefont {M.}~\bibnamefont {Bichler}}, \bibinfo {author}
  {\bibfnamefont {R.}~\bibnamefont {Meyer}}, \bibinfo {author} {\bibfnamefont
  {M.-C.}\ \bibnamefont {Amann}}, \ and\ \bibinfo {author} {\bibfnamefont
  {J.~J.}\ \bibnamefont {Finley}},\ }\bibfield  {title} {\enquote {\bibinfo
  {title} {Highly efficient single-photon emission from single quantum dots
  within a two-dimensional photonic band-gap},}\ }\href@noop {} {\bibfield
  {journal} {\bibinfo  {journal} {Phys. Rev. B},\ }\textbf {\bibinfo {volume}
  {77}},\ \bibinfo {pages} {073312} (\bibinfo {year}
  {2008}{\natexlab{b}})}\BibitemShut {NoStop}%
\end{thebibliography}

\providecommand{\noopsort}[1]{}\providecommand{\singleletter}[1]{#1}%

\end{document}